\documentclass[preprint,12pt,authoryear]{elsarticle}

\usepackage{amssymb}

\usepackage{graphicx}
\usepackage{subcaption}
\usepackage[section]{placeins}

\journal{Planetary and Space Science}

\begin{document}

\begin{frontmatter}



\title{Long-term Orbit Stability of the Apollo 11 “Eagle” Lunar Module Ascent Stage}


\author{James Meador}
\ead{james.meador@alumni.caltech.edu}

\affiliation{
			city={Mountain View},
            state={California},
            country={USA}}

\begin{abstract}
The Apollo 11 “Eagle” Lunar Module ascent stage was abandoned in lunar orbit after the historic landing in 1969. Its fate is unknown. Numerical analysis described here provides evidence that this object might have remained in lunar orbit to the present day. The simulations show a periodic variation in eccentricity of the orbit, correlated to the selenographic longitude of the apsidal line. The rate of apsidal precession is correlated to eccentricity. These two factors appear to interact to stabilize the orbit over the long term. 

\end{abstract}




\end{frontmatter}


\section{Introduction}
The non-uniform distribution of mass within the Earth’s Moon leads to unstable lunar orbits in most cases \nocite{Bel06}(Bell, 2006). Mass concentrations (“mascons”) cause a fairly rapid increase of eccentricity, resulting in surface impact. In 2012 the GRAIL mission, consisting of two identical satellites flying in formation, enabled mapping of the Moon’s gravitational field \nocite{Sid19} (Siddiqi 2019) with great precision, and led to the development of a series of lunar gravity models with successively higher fidelity \nocite{Lem13} \nocite{Lem14} \nocite{Goo16}(Lemoine et al., 2013; Lemoine, et al., 2014; Goossens et al., 2016).

The work described here was initiated with the hope of constraining the final impact location of the ascent stage of the Apollo 11 “Eagle”, which was jettisoned into a retrograde, near-equatorial lunar orbit after the historic landing in July, 1969. The Eagle was last tracked in a nearly circular orbit 99 by 117 km above the mean lunar radius. The fate of the spacecraft remains unknown, but it is widely assumed that it struck the surface decades ago \nocite{EdB20} (NASA, 2020). The author set about to simulate the orbit, using a GRAIL gravity model, assuming the orbit would destabilize quickly enough to constrain the search for the impact crater location. Instead, the simulated orbit exhibits long term quasi-periodic stability. Leaving aside the possibility for self-destruction as it aged, or propulsive events driven by leaks of its remaining fuel, according to these simulations this machine would still be in orbit today. The properties exhibited by the simulated orbit show evidence for a feedback mechanism that constrains it within a stable band.

\section{Methodology}
The simulator used is the General Mission Analysis Tool (GMAT) which was developed by NASA and is freely available online \nocite{NAS19}(NASA, 2019). This environment has been certified by NASA for use in mission planning. GMAT allows for the substitution of gravity models, and can natively load and interpret GRAIL models, which are also freely available. Lunar gravity is modeled using spherical harmonics, and GRAIL models are available with harmonic degree and order as high as 1200. The simulations reported here used the “gggrx\_1200a\_sha.tab” model. The computation required to propagate a simulation increases roughly as the square of the harmonic degree/order, and this places a practical limit on the fidelity of the model that can be used. Simulations described here were run with degree/order of the models set to 200, and in this case a simulation of ten years of spacecraft time completes in about 8 hours. Trial runs with degree and order as high as 1000 showed very similar results relative to “standard” runs at 200, suggesting that the major conclusions of this work would stand up even if significantly greater computation had been dedicated to the task. The simulation results also have been verified by a third party using a lunar orbit simulation tool that is completely independent of GMAT.

The nominal initial spacecraft orbit state comes primarily from the NASA Mission Report produced soon after the mission ended \nocite{NAS69}(NASA, 1969). Table 7-II from that report lists trajectory parameters for various events of the flight, including the moment when the Eagle was jettisoned. The parameters are expressed as time, latitude, longitude, altitude, velocity, flight path angle, and heading angle, relative to the Moon. These parameters map most directly to the “Planetodetic” state type in GMAT. The velocity must be adjusted to account for the rotating coordinate frame of this state type. Slight differences in flight path and heading angles due to the rotating coordinate frame were ignored. The altitude is expressed relative to the landing site, which is taken to be 1.9295 km below the mean lunar radius of 1738 km. 

\begin{figure}[h!]
	\centering
	\includegraphics[width=\linewidth]{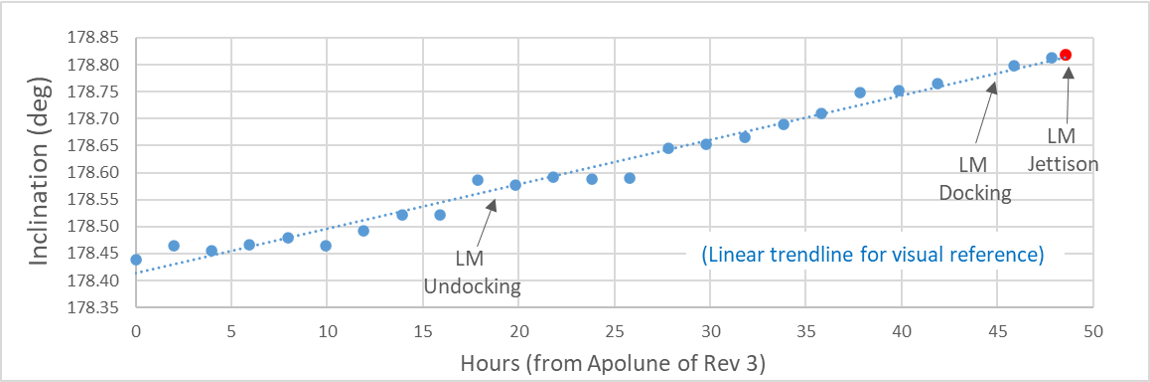}
	\caption{Plot of inclination values (from Murphy 1970) used to estimate the heading at jettison}
	\label{fig:Fig1}
\end{figure}

The heading angle listed in the Mission Report is incorrect, for unknown reasons. All early Apollo lunar orbits had retrograde, near-equatorial inclination, so the heading angle should have been close to -90 degrees. The listed value of -97.81° is inconsistent with any other known facts about the mission. To correct this presumably erroneous value, another source was used, which includes tables of Keplerian orbital parameters for the lunar phase of Apollo 11 \nocite{Mur70} (Murphy, 1970). Using inclination values from this table, extrapolating to the moment of jettison gives an inclination of around 178.82°. This data is shown in Figure \ref{fig:Fig1}.



This inclination translates to a heading angle of -89.63 in the planetodetic state at the moment of jettison. This value agrees well with other information about the trajectory at that time.

%

\newcommand{\centered}[1]{\begin{tabular}{l} #1 \end{tabular}}
\newcommand{\centeredc}[1]{\begin{tabular}{c} #1 \end{tabular}}

\begin{table}[htbp]
	\begin{center}
		\footnotesize
		\begin{tabular}{@{}l@{}|@{}c@{}|@{}c@{}|@{}l@{}}
			\hline
			\centered{ \textbf{Parameter} } &
			\centeredc{
				\textbf{Mission} \\
				\textbf{Report}  \\
				\textbf{Table 7-II}} &
			\centeredc{
				\textbf{GMAT} \\
				\textbf{Planetodetic}  \\
				\textbf{Value}} &
			\centered{
				\textbf{Notes} \\} \\
			\hline
			\centered{
				Time of \\
				Jettison } &
			\centeredc{
				130:09:31.2 \\
				(M.E.T.) \\ } &
			\centeredc{
				21 Jul 1969 \\
				23:41:31.2 \\ } &
			\centered{
				Mission Elapsed Time starts 16 Jul \\
				1969 at 13:32:00.00 UTC \\ } \\
			\hline
			\centered{
				Latitude } &
			\centeredc{1.10 °N } &
			\centeredc{1.10 °N } &
			\centered{ } \\
			\hline
			\centered{
				Longitude } &
			\centeredc{41.85 °E } &
			\centeredc{41.85 °E } &
			\centered{ } \\
			\hline
			\centered{
				Altitude \\
				(RMAG) } &
			\centeredc{61.6 n.mi. } &
			\centeredc{1850.15 km } &
			\centered{
				Mission altitude referenced to landing\\
				site, 1929 m below mean radius. } \\
			\hline
			\centered{
				Velocity \\
				(VMAG) } &
			\centeredc{5225.9 f.p.s. } &
			\centeredc{1.6313 km/sec } &
			\centered{
				Mission value is space-fixed.\\
				GMAT value is body-fixed. } \\
			\hline
			\centered{
				Flight Path \\
				Angle } &
			\centeredc{0.15 deg} &
			\centeredc{0.15 deg} &
			\centered{
				Mission value is space-fixed. Minor\\
				change for body-fixed is ignored. } \\
			\hline
			\centered{
				Heading \\
				Angle } &
			\centeredc{-97.81 deg} &
			\centeredc{-89.63 deg} &
			\centered{
				Mission value is erroneous. GMAT\\
				value derived from Murphy (1970). } \\
			\hline
		\end{tabular}				
		\caption{Nominal initial orbit state as derived from sources}
		\label{tab:Tab1}		
	\end{center}	
\end{table}

Table \ref{tab:Tab1} shows the nominal initial GMAT Planetodetic orbit state values, as derived from the report and paper. There are several sources of uncertainty in these values. The precision of the lunar ephemeris has improved since the time of the mission, so there is a problem to map historical descriptions of orbit states to the modern coordinate system. GMAT defaults to the DE405 ephemeris, and selenographic coordinates are referenced to the Moon\_PA system. Both the ephemeris and coordinates have been refined since 1969, but the exact changes are unknown; the early systems used magnetic tape and were not archived. Another source of uncertainty is due to the limits of the tracking and modeling used during the mission to estimate the spacecraft state. Still another uncertainty is whether the orbit parameters in the report are referring to the LM stage, or to the Command Module. At jettison the stage was moving radially at about 2 feet per second relative to the CM, representing an uncertainty of 0.02° in the flight path angle. 

To account for these unknowns, a Monte Carlo method was used to generate parameter sets that bracket the uncertainties. Most parameters were varied over a range of +/- 5\%. Velocity was varied over a range of +/- 0.2\% representing an error of +/- 10 feet/sec. This is generous, in that NASA documents from the time state that the velocity was accurate to within 0.5 f.p.s. The azimuth angle was varied over a range of +/- 0.25°, equivalent to a variation in the inclination of +/- 0.075°, roughly double the variation in the source table. Each parameter was varied randomly and uniformly over the given range, and then a secondary constraint was applied, to require that the perilune/apolune altitude values for the resulting orbit were within +/- 5\% of the nominal case. The range of values simulated is shown in Table \ref{tab:Tab2}.


\begin{table}[htbp]
	\begin{center}
		\footnotesize
		\begin{tabular*}{0.68\textwidth}{l|c|c|c}
			\hline						
			\textbf{Parameter}	&	\textbf{Nominal}	&	\textbf{Maximum}		&	\textbf{Minimum}	\\
			\hline
			Latitude (°N)		&	1.11				&	1.17				&	1.06	\\
			\hline
			Longitude (°E)		&	41.85				&	43.94				&	39.76	\\
			\hline
			RMAG (km)			&	1850.15				&	1855.76				&	1844.54	\\
			\hline
			VMAG (km/sec)		&	1.6313				&	1.6346				&	1.6280	\\
			\hline
			AZI	(deg)				&	-89.63				&	-89.37				&	-89.90	\\
			\hline
			HFPA (deg)				&	0.150				&	0.158				&	0.143	\\
			\hline
			Perilune (km)			&	101.77				&	106.86				&	96.68	\\
			\hline
			Apolune	(km)			&	115.27				&	121.03				&	109.51	\\
			\hline			
		\end{tabular*}		
		\caption{Range of values used for randomized trials}
		\label{tab:Tab2}		
	\end{center}
\end{table}

One hundred randomized parameter sets were each used to simulate one year in orbit, and all results showed similar behavior. Generally those parameter sets which started higher remained higher, and those that started lower remained lower. The 10 parameter sets with the lowest apolune during year 1 were simulated out to 10 years, and none show any sign of impact. The parameter set with the closest approach to the surface in the first year was simulated to 51 years, and no secular trend was observed in the minimum altitude. GMAT scripts, results, and other files are publicly available on GitHub \nocite{Mea20}(Meador, 2020).

A key element of the GMAT simulations is the “Propagator” object, which integrates the forces acting on the spacecraft in a stepwise fashion to estimate an updated state. The integrator used was “PrinceDormand78”, which is part of the standard GMAT package. In addition to the primary lunar gravity model, point masses were included for the Earth, Sun, and all other planets except Mercury. 

Solar radiation pressure was also incorporated in the simulations.  The spacecraft was modeled as a sphere with cross sectional area of 5 square meters. The spherical model is reasonable because the spacecraft attitude is not controlled, and the average surface presented towards the sun is likely to be smeared symmetrically over time. Two simulations were run for 10-year propagations with and without the effects of SRP. For the case with SRP, the area was increased to 10 sqm. The greatest difference in perilune altitude, during revolution 43301, is 0.38 km. (At 78.58 km with SRP versus 78.96 km without). The minimum perilune altitude across all revolutions is 16.04 km with SRP versus 16.07 km without, and these occur within 7 minutes of each other during revolution 33685, 7.6 years after jettison. The mean of the perilune altitude difference across all revolutions is less than one meter, showing that there is no secular trend induced by SRP. These results support the assertion that further refinement of the SRP model would not affect the principal result, that the orbits are long-lived.

\section{Results}

Figure 2 shows the simulated perilune altitude of the Eagle for the first year after it was jettisoned, and for the year ending on August 1, 2020. The eccentricity exhibits a cyclical behavior, with a fundamental period of around 24 days. (Perilune altitude and eccentricity are discussed interchangeably in this analysis. Lower perilune altitude relates linearly to higher eccentricity, as variations of the semi-major axis are minor.) The 24-day cycle is a key feature of the orbit, and it persists across decades. A long-periodic variation is also apparent, with minimum altitudes varying higher and lower every 4-5 cycles. Through selective alignment, the simulated behavior in 2020 can be matched up closely to that of 1969; the orbital behavior is stable over decades.

\begin{figure}[htbp]
	\centering	
	\begin{subfigure}[b]{\linewidth}
		\includegraphics[width=\linewidth]{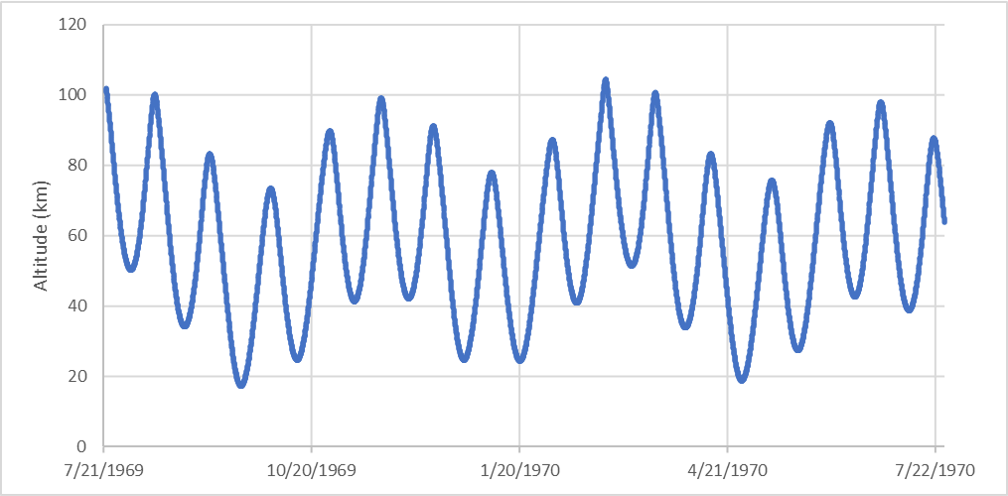}
	\end{subfigure}

	\begin{subfigure}[b]{\linewidth}
		\includegraphics[width=\linewidth]{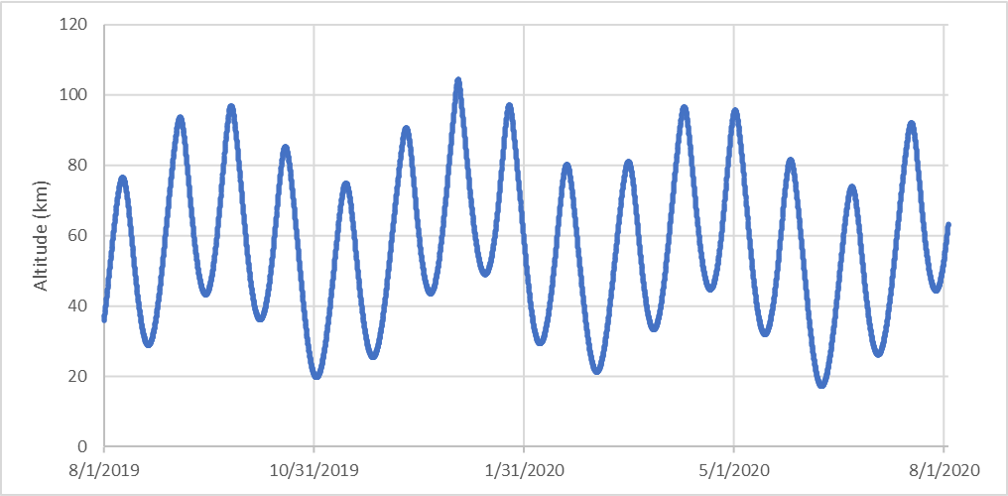}
	\end{subfigure}
	\caption{Simulated perilune altitude of the Eagle in 1969 and 2020. Altitude is above mean lunar radius.}
	\label{fig:Fig2}
\end{figure}

Figure 3 shows one year from jettison for the 100 randomized parameter sets. All 100 sets follow a similar pattern of cyclical eccentricity. Those parameter sets resulting in lower initial orbits generally maintain their rank throughout the year. Among all sets, the minimum altitude for the year is 6.3 km. (As described later, the minimum altitudes always occur in an area where lunar terrain is below 1.8 km.)

\begin{figure}[htbp]
	\centering
	\includegraphics[width=\linewidth]{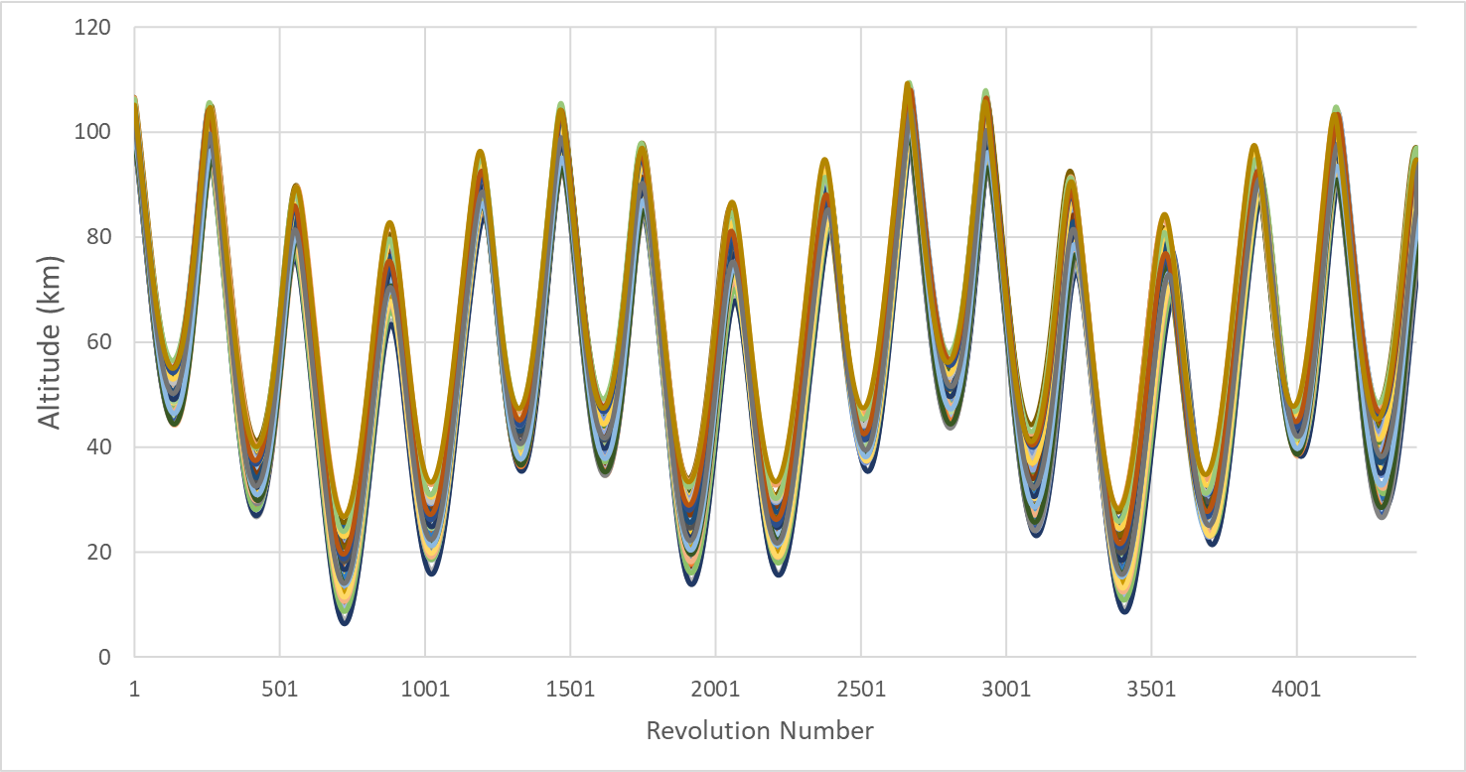}
	\caption{Perilune altitude for 100 variations of the nominal orbit, first year after jettison.}
	\label{fig:Fig3}
\end{figure}

Out of these 100 sets, the 10 with the lowest altitude during the first year were simulated to 10 years. All remained in orbit at the end of the period. Figure 4 shows the results for the tenth year. Although the phases of the cycles for each set have dispersed, all are generally following a pattern similar to the nominal case, with primary cycles of roughly 24 days and slower modulations. 

\begin{figure}[htbp]
	\centering
	\includegraphics[width=\linewidth]{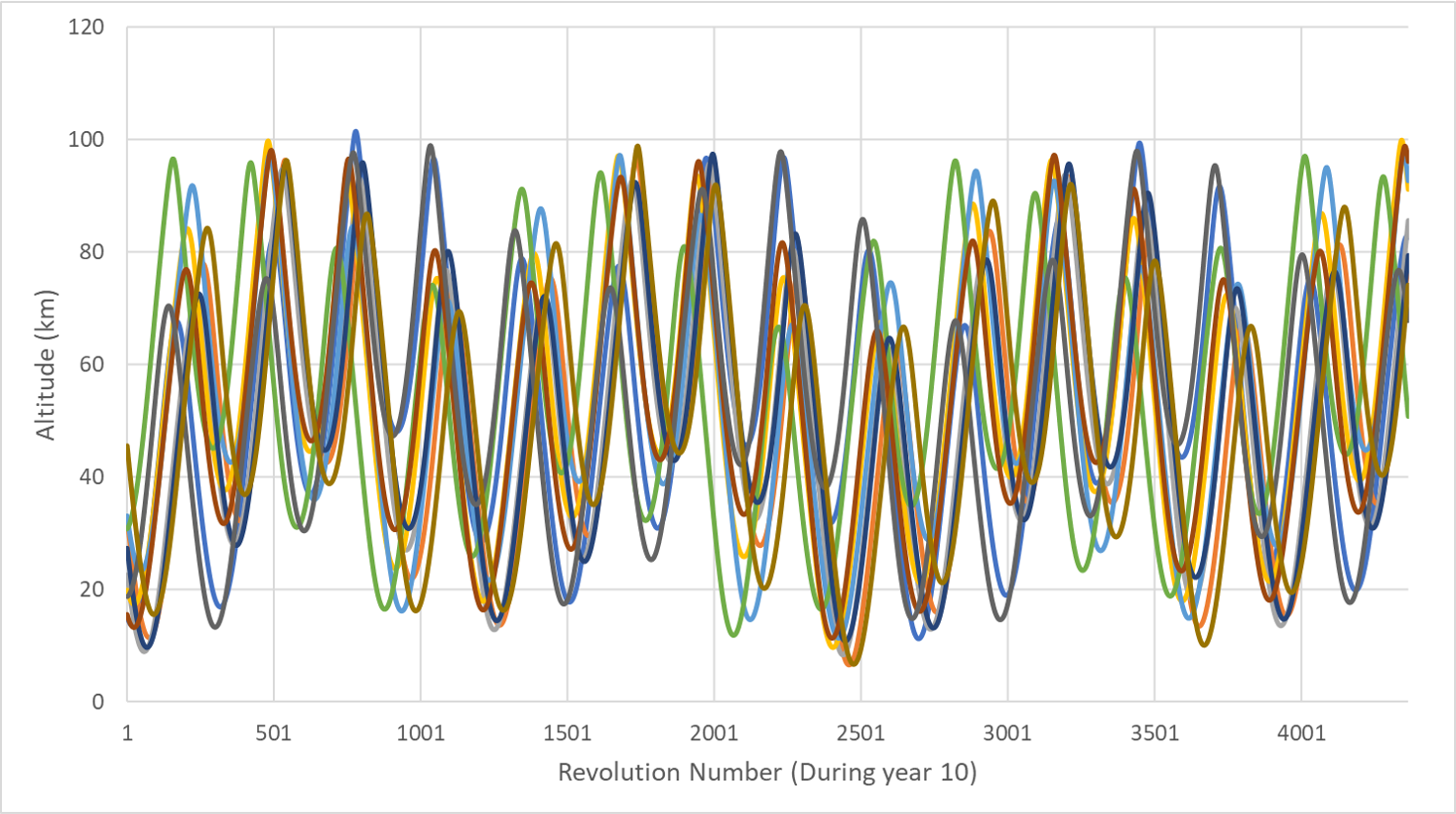}
	\caption{Perilune altitude ten years after jettison for the 10 parameter sets with lowest altitude during year one.}
	\label{fig:Fig4}
\end{figure}

The parameter set with the closest approach to the surface was simulated to 51 years. As can be seen in Figure 5, the same basic pattern persists across the full time period, and there is no apparent trend towards surface impact.

\begin{figure}[htbp]
	\centering
	\includegraphics[width=\linewidth]{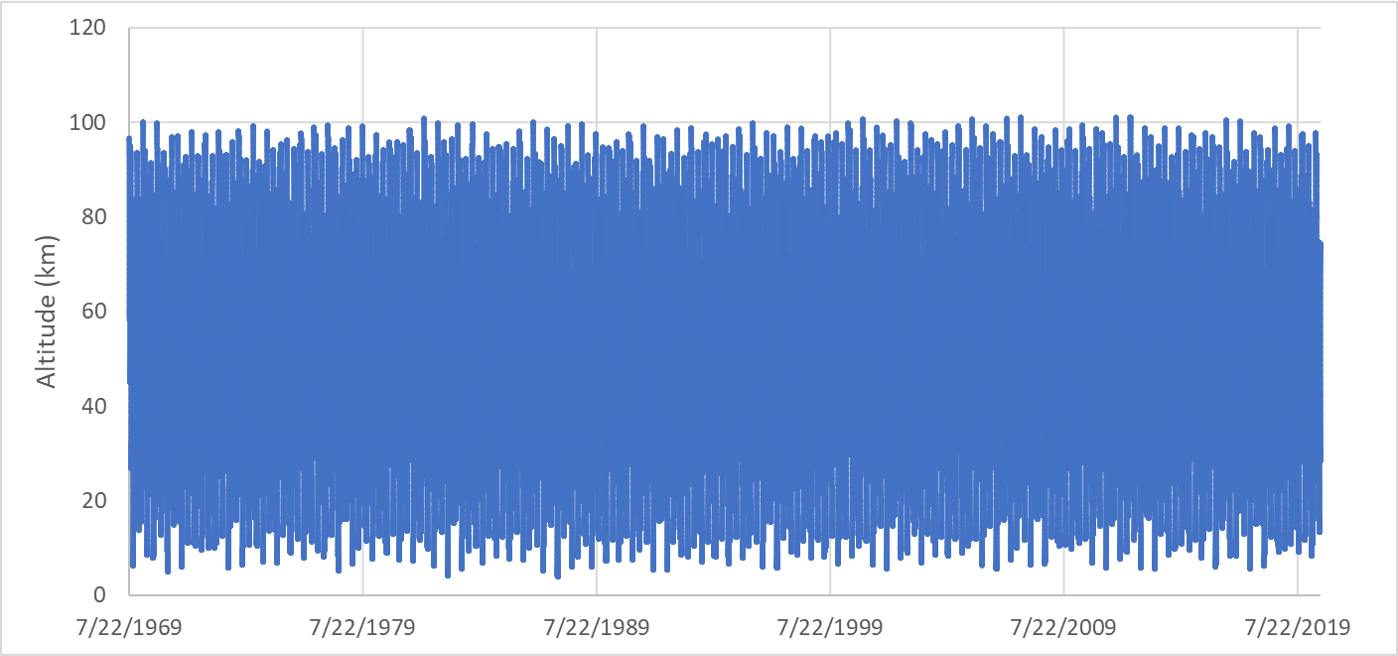}
	\caption{Perilune altitude for the parameter set with the lowest initial orbit, across 51 years.}
	\label{fig:Fig5}
\end{figure}

These numerical experiments support the hypothesis that even with the uncertainty of the initial conditions, the true orbit of the Eagle exhibits long term stability, and the spacecraft would not have impacted the Moon due to gravitational effects.

\section{Discussion}

The 24-day eccentricity cycles appear to be driven by equatorial mass concentrations. By modeling lunar gravity as a point source, other perturbations are seen to drive an annual variation of 2 km, versus the 80+ km variation seen in Figure \ref{fig:Fig5}.
Further evidence that mascons drive the cycles is apparent when the perilune altitude is plotted against selenographic longitude. Figure 6 shows altitude versus longitude across a 51-year simulation of the nominal orbit. (Most individual data points are not distinguishable.) There is obviously a strong correlation. Across decades, and hundreds of cycles, maximum eccentricity is reached when the perilune is on the near side of the Moon near longitude 30° East.

\begin{figure}[htbp]
	\centering
	\includegraphics[width=\linewidth]{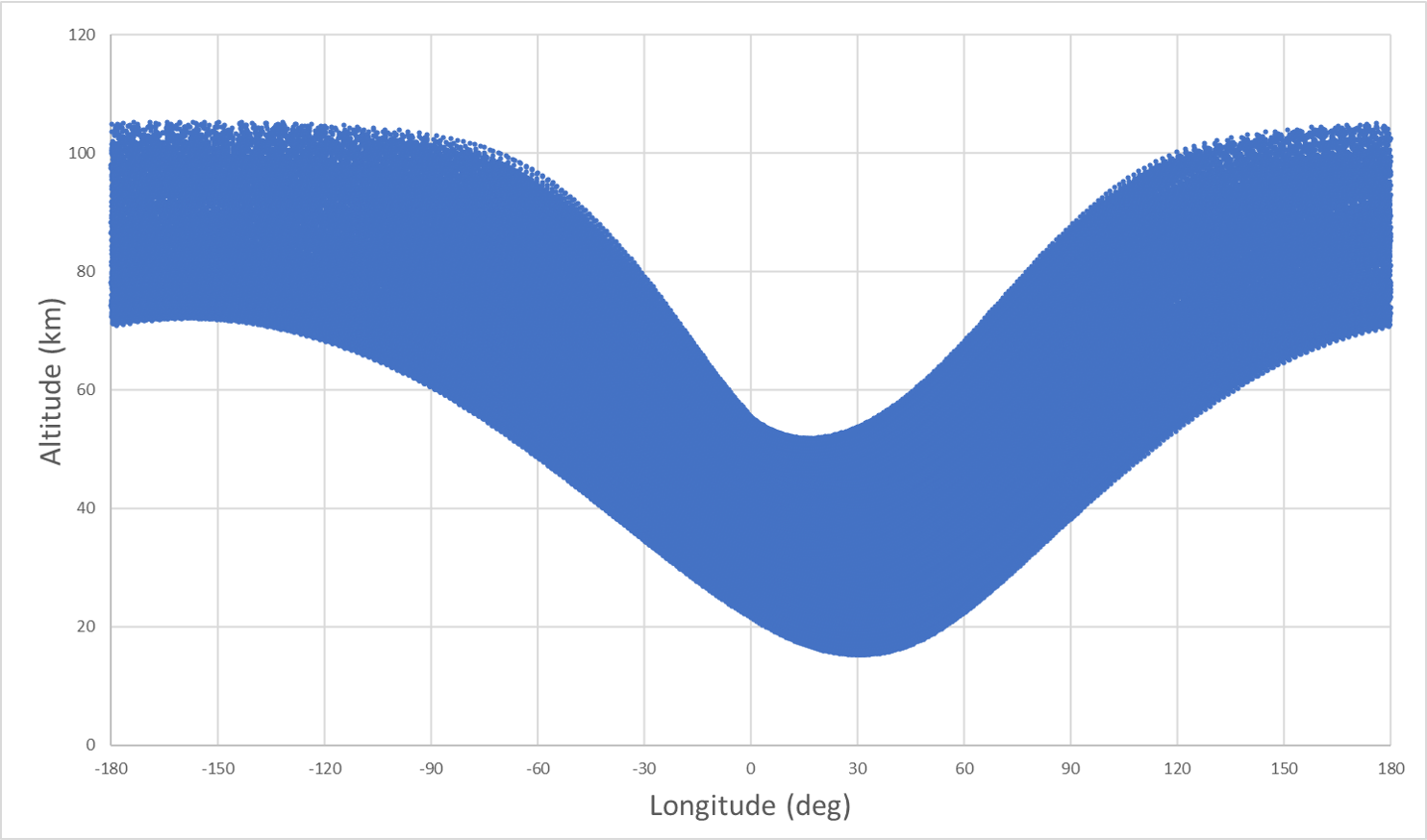}
	\caption{Perilune altitude vs. longitude, nominal orbit, 51 Years.}
	\label{fig:Fig6}
\end{figure}

The 24-day cycles align with the 27.32-day length of the sidereal month due to apsidial precession. The Eagle’s orbit is precessing in a direction opposite the rotation of the moon, so that the line of apsides of the orbit circles the moon a few days before the moon completes a full rotation around its axis. The perilune point moves a mean of 15°/day relative to selenographic longitude. Since the moon’s rotation is 13.17°/day, apsidal precession accounts for the additional 1.83°/day on average. 

The apsidal precession is not constant. Figure 7 shows that the rate of change of the perilune longitudinal point is faster when eccentricity is lower and perilune altitude is higher.

\begin{figure}[htbp]
	\centering	
	\begin{subfigure}[b]{0.5\linewidth}
		\includegraphics[width=\linewidth]{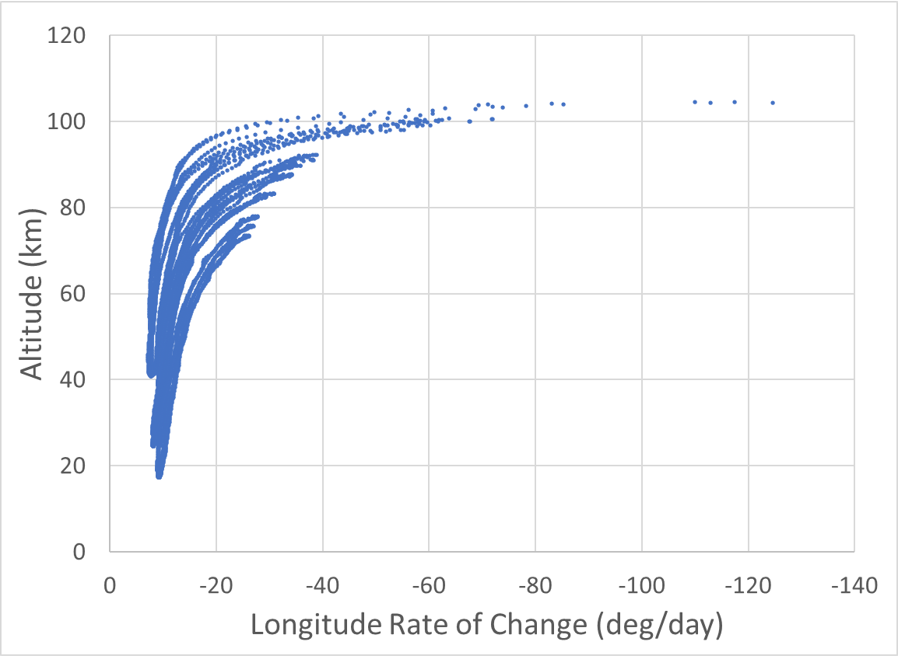}
	\end{subfigure}
	\begin{subfigure}[b]{0.47\linewidth}
		\includegraphics[width=\linewidth]{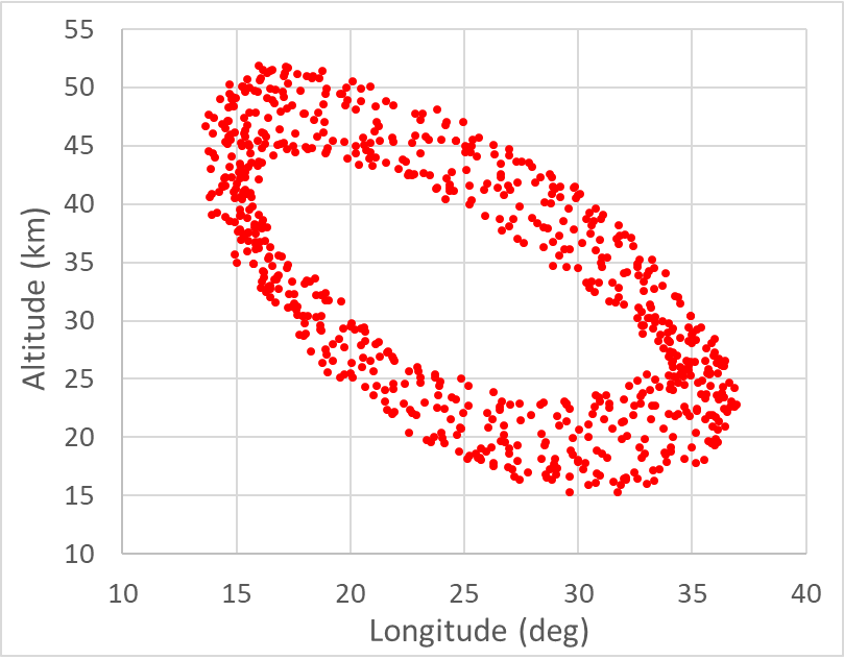}
	\end{subfigure}
	\caption{The left plot shows longitudinal rate of change vs. perilune altitude during the first year for the nominal orbit. The right plot shows the minimum altitude (per cycle) vs. longitude across 51 years.}
	\label{fig:Fig7}
\end{figure}

On the right in Figure 7 is a plot of the minimum perilune altitude of each 24-day cycle versus the selenographic longitude where that peak occurs. (This is a subset of the points plotted in Figure \ref{fig:Fig6}.) Successive points are counter-clockwise around the torus, with 4 to 5 points to complete one long cycle. This toroidal pattern ties back to the long-periodic modulation of the 24-day cycles seen in Figure 1. During cycles of higher eccentricity, the perilune minima occur between 25° and 35° East. During cycles with lower eccentricity, the minima occur further to the West.

As a final observation, one can measure the period of each cycle, defined as the time required for perilune longitude to circle the Moon, and plot this along with the minimum perilune altitude for that cycle, as in Figure 8. Clearly these quantities change in opposition to each other. The dependency of eccentricity on longitude, coupled with the dependency of apsidal precession rate on eccentricity, appear to have the combined effect of maintaining the orbit in a quasi-stable, wobulating state indefinitely.

\begin{figure}[htbp]
	\centering
	\includegraphics[width=\linewidth]{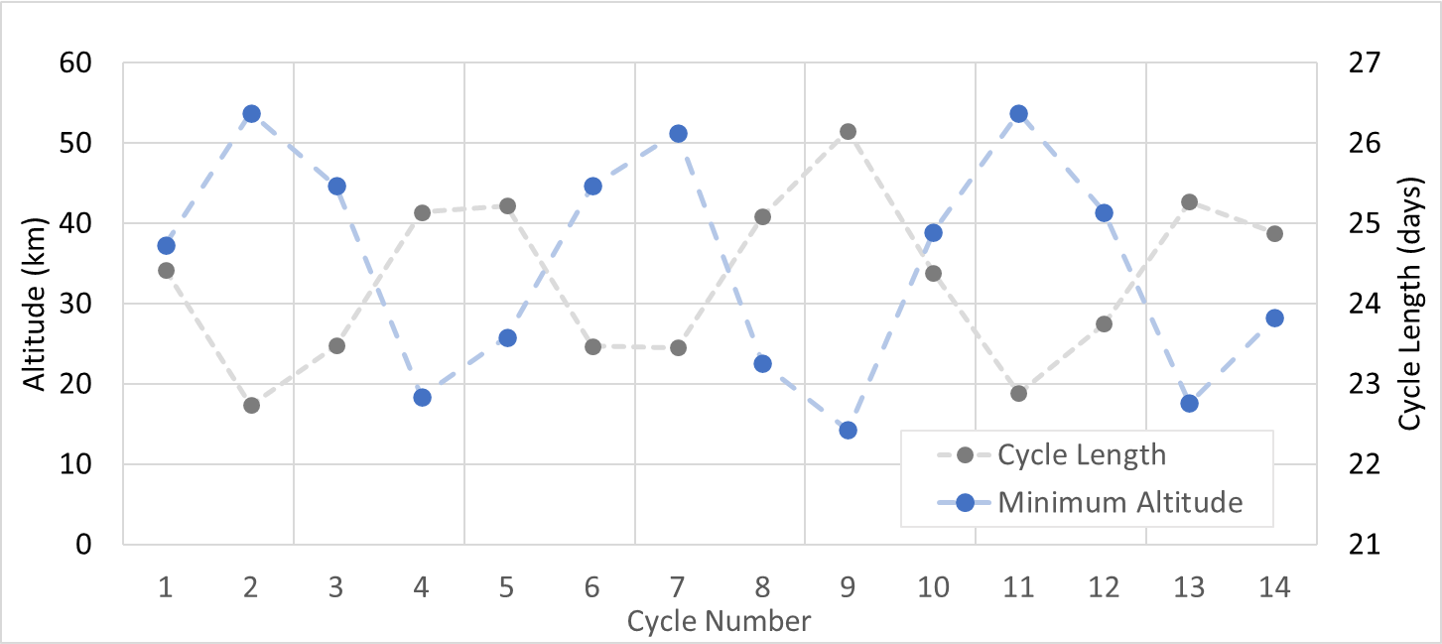}
	\caption{Showing cycle period and minimum cycle altitude for the first year of the nominal orbit.}
	\label{fig:Fig8}
\end{figure}

These results indicate that the “Eagle” might remain in lunar orbit today. It must be remembered that the Lunar Modules were designed for 10-day missions, and little consideration was given to long-term reliability. For this reason, fuel leaks might have resulted in propulsive events or even complete destruction at any time after the craft was jettisoned. Although catastrophic outcomes are possible, there exists some possibility that this machine might have reached an inert state, allowing it to remain in orbit to the present day. If so, it should be detectable by radar, similarly to the way that the Indian Moon orbiter Chandrayaan-1 was relocated in 2016 \nocite{JPL17}(JPL, 2017). The author hopes that even if this possibility is remote, given the importance of the artifact, that radar time could be allocated to a scan of the regions close to the lunar limb at near-equatorial latitudes. A rough analysis indicates that Eagle would be more than 125 km above the surface in about 25\% of limb crossings. If one assumes radar is able to detect objects at this altitude, then four judiciously chosen 2-hour observation periods should provide sufficient coverage to possibly relocate one of the most important artifacts in the history of space exploration.

\section{Acknowledgments}

The author is grateful to many professionals in the field who kindly responded to emails from a stranger with encouragement and suggestions, including Dr. Alex Konopliv, Laura Plice, Andres Dono Perez, Dr. Jack Burns, and Dr. Philip J. Stooke. The author is especially grateful to Michael Khan of the European Space Agency for his feedback and encouragement, and for his work to confirm the basic results using an independent simulation environment.




\bibliographystyle{elsarticle-harv} 
\bibliography{MyPaperLibrary}

\begin{thebibliography}{11}
\expandafter\ifx\csname natexlab\endcsname\relax\def\natexlab#1{#1}\fi
\providecommand{\url}[1]{\texttt{#1}}
\providecommand{\href}[2]{#2}
\providecommand{\path}[1]{#1}
\providecommand{\DOIprefix}{doi:}
\providecommand{\ArXivprefix}{arXiv:}
\providecommand{\URLprefix}{URL: }
\providecommand{\Pubmedprefix}{pmid:}
\providecommand{\doi}[1]{\href{http://dx.doi.org/#1}{\path{#1}}}
\providecommand{\Pubmed}[1]{\href{pmid:#1}{\path{#1}}}
\providecommand{\bibinfo}[2]{#2}
\ifx\xfnm\relax \def\xfnm[#1]{\unskip,\space#1}\fi
\bibitem[{Agle(2017)}]{JPL17}
\bibinfo{author}{Agle, D.C.}, \bibinfo{year}{2017}.
\newblock \bibinfo{title}{New nasa radar technique finds lost lunar
  spacecraft}.
\newblock \URLprefix \url{https://www.jpl.nasa.gov/news/news.php?feature=6769}.
\bibitem[{Bell()}]{Bel06}
\bibinfo{author}{Bell, T.},
  \bibinfo{year}{2013}.
\newblock \bibinfo{title}{Bizarre lunar orbits}.
\newblock \URLprefix \url{{https://science.nasa.gov/science-news/science-at-nasa/2006/06nov\_loworbit/}}.
\bibitem[{Goossens and et. al.(2016)}]{Goo16}
\bibinfo{author}{Goossens, S.}, \bibinfo{author}{et al.}, \bibinfo{year}{2016}.
\newblock \bibinfo{title}{A global degree and order 1200 model of the lunar
  gravity field using grail mission data}, \bibinfo{organization}{Lunar and
  Planetary Science Conference}. p. \bibinfo{pages}{Abstract 1484}.
\bibitem[{Lemoine and et. al.(2013)}]{Lem13}
\bibinfo{author}{Lemoine, F.G.}, \bibinfo{author}{et al.},
  \bibinfo{year}{2013}.
\newblock \bibinfo{title}{High‒degree gravity models from grail primary
  mission data}.
\newblock \bibinfo{journal}{Journal of Geophysical Research: Planets} ,
  \bibinfo{pages}{1676--1698}.
\bibitem[{Lemoine et~al.(2014)Lemoine, Goossens, Sabaka, Nicholas, Mazarico,
  Rowlands, Loomis, Chinn, Neumann, Smith and Zuber}]{Lem14}
\bibinfo{author}{Lemoine, F.G.}, \bibinfo{author}{et al.},
 \bibinfo{year}{2014}.
\newblock \bibinfo{title}{Grgm900c: A degree 900 lunar gravity model from grail
  primary and extended mission data}.
\newblock \bibinfo{journal}{Geophysical Research Letters} \bibinfo{volume}{41},
  \bibinfo{pages}{3382--3389}.
\newblock \DOIprefix\doi{https://doi.org/10.1002/2014GL060027}.
\bibitem[{Meador(2020)}]{Mea20}
\bibinfo{author}{Meador, J.}, \bibinfo{year}{2020}.
\newblock \bibinfo{title}{Simulation files}.
\newblock \URLprefix \url{https://github.com/RogerTwank/Eagle}.
\bibitem[{Murphy(1970)}]{Mur70}
\bibinfo{author}{Murphy, J.P.}, \bibinfo{year}{1970}.
\newblock \bibinfo{title}{Lunar Gravity Models for Improved Apollo Orbit
  Computation}.
\newblock \bibinfo{type}{Technical Report}.
\newblock \URLprefix \url{https://ntrs.nasa.gov/api/citations/19700028128/downloads/19700028128.pdf}.
\bibitem[{NASA(1969)}]{NAS69}
\bibinfo{author}{NASA}, \bibinfo{year}{1969}.
\newblock \bibinfo{title}{Apollo 11 mission reports}.
\newblock \URLprefix \url{https://www.hq.nasa.gov/alsj/a11/a11mr.html}.
\bibitem[{NASA(2019)}]{NAS19}
\bibinfo{author}{NASA}, \bibinfo{year}{2019}.
\newblock \bibinfo{title}{General mission analysis tool (gmat)}.
\newblock \URLprefix \url{https://opensource.gsfc.nasa.gov/projects/GMAT/index.php}.
\bibitem[{NASA(2020)}]{EdB20}
\bibinfo{author}{NASA}, \bibinfo{year}{2020}.
\newblock \bibinfo{title}{Apollo 11 lunar module / easep}.
\newblock \URLprefix \url{https://nssdc.gsfc.nasa.gov/nmc/spacecraft/display.action?id=1969-059C}.
\bibitem[{Siddiqi(2019)}]{Sid19}
\bibinfo{author}{Siddiqi, A.A.}, \bibinfo{year}{2019}.
\newblock \bibinfo{title}{Grail (ebb and flow)}.
\newblock \URLprefix \url{https://solarsystem.nasa.gov/missions/grail/in-depth/}.

\end{thebibliography}

%
%
%
\end{document}